\documentclass[12pt,a4paper]{article}
\usepackage{graphics}
\newcommand\smfrac[2]{{\textstyle{\frac{#1}{#2}}}}

\newcommand\ltap{\raisebox{-.45ex}{\rlap{$\sim$}} \raisebox{.45ex}{$<$}}

\begin{document}
\begin{titlepage}
\begin{flushright}
  RAL-TR-1998-067 \\
  IFUM-633-FT \\
  hep-ph/9809451
\end{flushright}
\par \vspace{10mm}
\begin{center}
{\Large \bf
Matrix Element Corrections to Parton Shower\\[1ex]
Simulations of Heavy Quark Decay}
\end{center}
\par \vspace{2mm}
\begin{center}
{\bf G. Corcella}\\
\vspace{2mm}
{Dipartimento di Fisica, Universit\`a di Milano
  and INFN, Sezione di Milano}\\
{Via Celoria 16, 20133 Milano, Italy}
\par \vspace{5mm}
and
\par \vspace{5mm}
{\bf M.H. Seymour}\\
\vspace{2mm}
{Rutherford Appleton Laboratory, Chilton,}\\
{Didcot, Oxfordshire.  OX11 0QX\@.  U.K.}
\end{center}

\par \vspace{2mm}
\begin{center} {\large \bf Abstract} \end{center}
\begin{quote}
\pretolerance 10000
Parton showers are accurate for soft and/or collinear emission, but for
a good description of the whole of phase space they need to be
supplemented by matrix element corrections.  In this paper, we discuss
matrix element corrections to the decay $t\to Wb$ and apply our results
to the HERWIG Monte Carlo event generator.  The phenomenological results
show marked improvement relative to previous versions and agree well
with the exact first-order matrix-element calculation.
\end{quote}

\vspace*{\fill}
\begin{flushleft}
  RAL-TR-1998-067 \\
  IFUM-633-FT \\
  September 1998
\end{flushleft}
\end{titlepage}

\section{Introduction}

Heavy quark physics is at present one of the main subjects of
theoretical and experimental particle physics[\ref{FMNR}] and one of the
objectives of future experiments at the LHC and NLC\@.  Although total
cross sections and other inclusive quantities can be reliably calculated
using fixed-order perturbative QCD, for analyses that require a detailed
description of final state properties one needs to sum large logarithmic
contributions to all orders and include non-perturbative hadronization
effects.  Monte Carlo event generators use parton shower models to
perform this resummation and phenomenologically-inspired models of the
hadronization process.

The parton shower approach relies on the universality of the QCD matrix
elements in the dominant regions of phase space: soft and/or collinear
emission[\ref{factorization}].  Multi-parton final states are built up
by starting from a few well-separated hard partons and sequentially
adding one extra parton at a time, distributed according to the soft and
collinear factorization formulae.  In the collinear limit it is
straightforward to formulate this as a probabilistic Markov chain, as
the factorization formulae are directly written in terms of probability
distributions.  In the soft limit, the factorization theorem applies at
the amplitude level, and many interfering amplitudes contribute equally,
so it would appear that no probabilistic approach could be formulated.
However, the remarkable result[\ref{MW1}] is that the interference is
entirely destructive outside angular-ordered regions of phase space.
Therefore parton shower algorithms can be extended to correctly treat
both regions, simply by using opening angle as the ordering variable
within the collinear approach.  This is the basis of coherence-improved
parton shower algorithms such as HERWIG[\ref{HERWIG}].

Coherence is also important in setting the initial conditions of the
parton shower: each hard parton may only radiate into an angular region
extending as far as its `colour partner'[\ref{MW2}].  The assignment of
colour partners is controlled by the hard matrix element, and thus the
properties of the parton shower and the resulting jets are influenced by
the kinematics of the hard scattering, giving a perturbative origin to
coherent phenomena like the string effect[\ref{ADKT}].

It is straightforward to incorporate quark masses into a
coherence-improved parton shower algorithm[\ref{MW3}].  In heavy quark
production, the mass terms result in an angle-dependent suppression that
can be approximated by an angular cutoff: the massive quark radiates
like a massless one at large angles, and not at all at small angles,
$\theta \ltap m_q/E_q$.  Heavy quark decay acts like a new hard process
in which gluons are emitted coherently by the heavy quark and by its
lighter quark decay product.  In principle, this separation into
production and decay phases is only valid for gluon energies above the
decay width of the quark, which for top quarks is of order 1~GeV, and
below that the coherence of the whole production and decay chain should
be taken into account[\ref{KS}].  In practice however, 1~GeV is close
enough to the cutoff that terminates the parton shower that these
effects may be neglected in the parton shower and only need to be
incorporated in the non-perturbative hadronization phase.

Although the soft and collinear regions dominate the emission
probabilities, many experimental observables are most sensitive to hard
non-collinear emission.  A poignant example is the kinematic
reconstruction of the mass of decaying top quarks, in which the
treatment of additional jets can make a sizeable difference, leading to
an uncertainty in the final result that is almost as large as the
experimental uncertainty[\ref{top}].  Unfortunately, there is no
guarantee that a parton shower algorithm will describe such emission
well.  It should, however, be well-described by the next-order (but
still tree-level) matrix element.  Thus, if one wants to have a good
description of the bulk properties of events and the rare hard
wide-angle emissions simultaneously, one must combine the parton shower
and matrix element approaches to provide a process-specific `matrix
element correction' to the parton shower.

In general there are two shortcomings to be corrected: it is possible
that owing to the angular-ordering condition, there are regions of phase
space that are not covered at all by the parton shower (which we call
`dead zones'); and it is possible that even in the region it does cover,
the parton shower is not a good approximation to the exact matrix
element.  General solutions to both these problems (which we call the
`hard' and `soft' corrections respectively) were described in
Ref.~[\ref{S1}], and have been incorporated into HERWIG for $e^+e^-$
annihilation[\ref{S2}] and DIS[\ref{S3}].  Similar issues were discussed
for the JETSET/PYTHIA parton shower algorithm in Ref.~[\ref{AS}].

In this paper we discuss the implementation of matrix element
corrections to heavy quark decay in HERWIG and show first results.  In
section~2 we recall the relevant features of HERWIG's parton shower
algorithm.  In sections~3 and~4 we discuss the hard and soft matrix
element correction respectively.  In section~5 we present some
phenomenological distributions for top quark decay and compare them with
the exact QCD results and those obtained from the previous version of
HERWIG\@.  Finally in section~6 we make some concluding remarks.

\section{The parton shower algorithm}

For the decay\footnote{In fact we use exactly the same method for all
  heavy (bottom and above) quarks, but for clarity we always use the
  language of top quark decay.  Note also that although we describe the
  decay as $t\to bW(g)$, we actually use the full three-body matrix
  elements for $t\to bf\bar{f}'$ so that the $f\bar{f}'$ mass
  distribution and angular correlations are correct.} $t\to bW\!$, the
elementary probability to emit an additional gluon is given by
\begin{equation}
  dP={{d\Gamma (t\to bW\!g)}\over{\Gamma_0}},
\end{equation}
where $\Gamma_0$ is the width for the Born process.  In the soft and
collinear limits, this can be approximated by the universal elementary
probability for the emission of an additional parton in any hard
process:
\begin{equation}
  \label{elementary}
  dP={{dq^2}\over{q^2}}\;{{\alpha_S}\over {2\pi}}\;P(z)dz\;
  {{\Delta_S(q^2_{\mathrm{max}},q^2_c)}\over{\Delta_S(q^2,q^2_c)}}\ .
\end{equation}
The HERWIG parton shower is ordered according to the variable
$q^2=E^2\xi$, where $E$ is the energy of the emitter, $\xi={{q_1\cdot
    q_2}\over{E_1 E_2}}$ and $q_{1,2}$ being the four-momenta of the
produced partons; $z$ is the energy fraction of the emitted parton in
the showering frame.  For massless partons we have $\xi=1-\cos\theta$,
where $\theta$ is the opening angle of emission, so that ordering in
$q^2$ corresponds to ordering in angles.

The crucial quantity $\Delta_S(q^2,q^2_c)$ is the Sudakov form factor,
representing the probability that no {\em resolvable\/} radiation is
emitted from a parton whose upper limit on emission is~$q^2$.  The
resolution requirement, which in the case of HERWIG is a cutoff on
transverse momentum, is $q_c^2$.  Thus the ratio of form factors
appearing in Eq.~(\ref{elementary}) represents the probability that the
emission considered is the first i.e.\ the highest in $q^2$.  In
diagrammatic language, this actually sums up the contributions of
virtual and unresolvable real emission to all orders.

Using HERWIG's definitions for $q^2$ and $z$, the results of the parton
shower are not Lorentz covariant, although they become so in the exactly
soft or collinear limits in which the algorithm is formally valid.
Nevertheless, the extrapolations away from those limits depend on the
frame in which the shower is developed.  HERWIG uses a different frame
for each jet $i$, in which $\xi_{i\mathrm{max}}=1$ and hence, since
$q^2=E^2\xi$, $E_i=q_{i\mathrm{max}}$.  After parton showering, the
momenta of the jets cannot be identical to the momenta of the partons
that initiated them, since the latter are on mass-shell, so some
momentum must be transferred between jets to conserved energy-momentum
overall.  The precise details become irrelevant in the soft and
collinear limits in which the algorithm is valid, but are again
important for the extrapolation away from that limit that we are
concerned with here.  The procedure is again frame-dependent and is
performed by Lorentz boosting each jet along its direction in the
rest-frame of the hard process that produced it.

Colour coherence in the hard process dictates the value of
$q^2_{\mathrm{max}}$, which for colour-connected partons $i$ and $j$ are
related by $q_{i\mathrm{max}}q_{j\mathrm{max}}=p_i\cdot p_j$.  In
principle, this is the only requirement, and one has a free choice for
the individual values $q_{i\mathrm{max}}$ and $q_{j\mathrm{max}}$,
corresponding to different choices of frame, but in most cases,
symmetric limits are chosen:
$q_{i\mathrm{max}}^2=q_{j\mathrm{max}}^2=p_i\cdot p_j$.  However, for
the special case of top quark decay, we choose
$E_t=q_{t\mathrm{max}}=m_t$ and therefore
$E_b=q_{b\mathrm{max}}={{p_t\cdot p_b}\over{m_t}}$, corresponding to the
top rest-frame, in which the top quark itself does not radiate.

Because of this choice, and the fact that HERWIG only radiates into the
angular-ordered region $\xi<1$, there is no radiation in the $W$
hemisphere, while in the full matrix element such radiation is
suppressed but not absent.  Therefore the dead zone actually includes
part of the soft singularity, in contrast to the $e^+e^-$ and DIS cases
in which soft emission from one or other jets covers the whole of the
angular range.  This poses some additional problems, which we describe
heuristically here, and in more detail in the following sections.

Since HERWIG's dead zone includes part of the soft singularity, the
total amount of emission into it, if calculated na\"\i vely, is
infinite.  Therefore the hard correction method of Ref.~[\ref{S1}]
cannot be used.  The correct solution to this problem is to modify
HERWIG's parton shower algorithm so that it populates the backward
direction.  Unfortunately, this is far from straightforward to do.
Instead, we look for simpler, more approximate, solutions.

If we apply a cutoff on the gluon energy then, provided the total
probability of emitting into the dead zone above the cutoff is small,
the hard correction can be used.  Although this is only an approximation
we can easily check, by varying the cutoff, how good it is.  In
section~5 we show that varying the cutoff in a reasonable range gives a
negligible effect on physical distributions.

However, the fact that we apply an energy cut in the hard correction has
implications for the soft correction.  The general idea of HERWIG's
approach to modelling emission from heavy quarks[\ref{MW3}] is that in
the soft limit, the emission pattern is similar to that from a light
quark at large angles, but is smoothly suppressed at small angles,
according to the replacement
\begin{equation}
  \frac1{\theta^2} \longrightarrow
  \frac{\theta^2}{(\theta^2+1/\gamma^2)^2},
\end{equation}
where $\gamma$ is the Lorentz factor of the emitting quark.  This is
approximated by a step function at $\theta=1/\gamma$.  While this does
not give a very good description of the angular distribution of emitted
gluons, it does give a good approximation to the total amount of gluon
emission, which is important for the total energy loss of the heavy
quark for example.  In other words, we get the amount of gluon emission
about right, but tend to put it in the wrong place.  For top decay, this
is at its most extreme, since we work in its rest-frame in which the
neglected angular region is the whole of the $W$ hemisphere.

If we were to apply a soft matrix element correction in the usual way,
we would decrease the amount of emission in the $b$ hemisphere but,
because we use a cutoff on the gluon energy in the hard correction, for
gluons below this cutoff there would be no corresponding increase in the
$W$ hemisphere.  We would therefore get the total amount of emission
wrong, and hence quantities like the $b$ quark's energy spectrum would
be poorly described.  To avoid this, we use the same cutoff in the soft
correction, and only correct the distribution of gluons with energy
above it.  Below the cutoff, the standard parton shower description is
used alone.

In summary, we can construct the matrix element corrections as the
combination of hard and soft corrections in exactly the usual way, but
only in the region of gluon energies above a given cutoff,
$E_{\mathrm{min}}$.  Below the cutoff, we use the parton shower
uncorrected.  We use 2~GeV as the default value of the cutoff.

Finally, as discussed in Ref.~[\ref{S1}], one should include a form
factor in the region covered by the hard correction, to ensure smooth
matching at the boundary.  Since we approach the soft singularity in the
top decay case, this could in principle be more important than in other
cases.  However, we find that even with a cutoff as small as 1~GeV, the
form factor is never more than a few percent below unity, so does not
significantly affect any physical distributions.  We therefore neglect
it for simplicity, as is already done in the $e^+e^-$ and DIS cases.

\section{Hard Corrections}

\begin{figure}
  \centerline{%
    \resizebox{!}{6cm}{\includegraphics{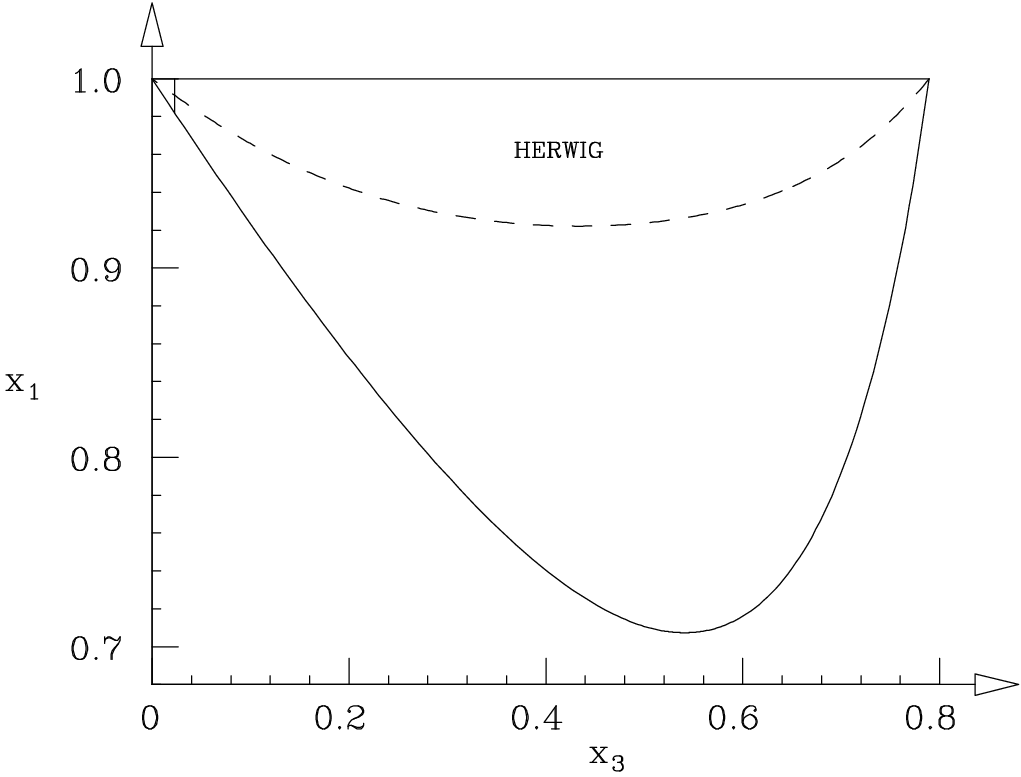}}
    }
  \caption[]{Phase space limits on
    $x_1={{2p_1 \cdot q}\over{m_t^2}}-a$ and
    $x_3={{2p_3 \cdot q}\over{m_t^2}}$ in the decay $t\to bW\!g$ (solid)
    and the edge of the region covered by HERWIG (dashed).  Also shown,
    barely visible in the top-left corner, is a 2~GeV cutoff on the
    gluon energy.}
  \label{ps}
\end{figure}

The main step in constructing both the hard and soft matrix element
corrections is to relate the variables generated by HERWIG, $z$ and
$\xi$, to the kinematic variables we use in the matrix element
calculation.  We parametrize the phase space of the process $t(q)\to
W(p_1) b(p_2) g(p_3)$ in terms of the variables\footnote{To simplify the
  formulae in this paper, we neglect the $b$ mass, although we do include
  it in HERWIG, and hence in all figures that we show.}
\begin{equation}
  x_1=1-{{2p_2\cdot p_3}\over {m_t^2}}=
  {{2p_1\cdot q}\over {m_t^2}}-a\ ,
\end{equation}
where we have defined\footnote{Recalling that we use the three-body
  matrix elements for $t\to bf\bar{f}'$, $a$ is actually defined to be
  $(p_f+p_{\bar{f}'})^2/m_t^2$, which varies from event to event, but
  this difference is unimportant here.}
$a=m_W^2/m_t^2$, and
\begin{equation}
  x_3={{2p_3\cdot q}\over {m_t^2}}\ .
\end{equation}
The phase space limits for the decay $t\to bWg$ are as follows:
\begin{eqnarray}
  \hspace{-2cm}
  \frac{a\,x_3}{1-x_3}+(1-x_3) \;<&x_1&<\; 1,
  \\\hspace{-2cm}
  0 \;<&x_3&<\; 1-a.
\end{eqnarray}
In order to express $x_1$ and $x_3$ in terms of $\xi$ and $z$, we
observe that the Lorentz boost of the $b$ quark's parton shower takes
place in the top rest frame, so that the mass, $m$, of the $b$--$g$ jet
and its transverse momentum, $q_t$, relative to the $b$-$W$ axis are
conserved.  We also define the energy of the b quark in the showering
frame to be $\sqrt{c}m_t$ with, because HERWIG uses the top rest frame,
$c=(1-a)^2/4$.  In terms of the showering variables, we obtain:
\begin{eqnarray}
  m^2   &=& 2z(1-z)\xi\;c\,m_t^2,\\
  q_t^2 &=& \frac{z^2(1-z)^2}{1-2z(1-z)\xi}\xi(2-\xi)\;c\,m_t^2,
\end{eqnarray}
and in terms of the matrix element variables:
\begin{eqnarray}
  m^2   &=& m_t^2(1-x_1),\\
  q_t^2 &=& m_t^2\frac{(1-x_1)(x_1+x_3(2-x_1-a)-x_3^2-1)}{(x_1+a)^2-4a}.
\end{eqnarray}
They can be combined to give:
\begin{eqnarray}
  z &=& \frac12\Biggl\{1+
  \frac{\sqrt{(x_1-(1-c))/c}}{\sqrt{(x_1+a)^2-4a}}(2x_3+x_1-2+a)\Biggr\},\\
  \xi &=& \frac{2(1-x_1)((x_1+a)^2-4a)}
  {(1-x_1)(2x_3+x_1-2+a)^2+4c(1-x_3)(x_1+x_3-1)-4a\,c\,x_3}.
\end{eqnarray}
The region not populated by HERWIG can be calculated from the condition
$\xi>1$:
\begin{eqnarray}
  0\;<&x_3&<\; 1-a,\phantom{+\frac{a\,x_3}{1-x_3}} \\
  (1-x_3)+\frac{a\,x_3}{1-x_3}\;<&x_1&<\;x_{1\mathrm{max}},
\end{eqnarray}
with $x_{1\mathrm{max}}$ being the solution of the cubic equation
\begin{eqnarray}
  x_1^3 + (3+2a-4x_3)x_1^2 + (a^2-4(1-x_3)(2-c-x_3)-2a(3+2x_3))x_1
\nonumber\\
  + a(4-a)(2-c)+(1-c)(2(1-x_3)-a)^2 &=& 0.
\end{eqnarray}
Such a solution can be expressed in the following form:
\begin{equation}
  x_{1\mathrm{max}}=2\Re\left\{\left(r+i\sqrt{q^3-r^2}\right)^{1/3}\right\}
  -{1\over 3}(3+2a-4x_3),
\end{equation}
where we have defined
\begin{equation}
  q = {1\over 9}(3+2a-4x_3)^2
  -{1\over 3}\left[a^2-4(1-x_3)(2-c-x_3)-2a(3+2x_3)\right],
\end{equation}
and
\begin{eqnarray}
  r &=& {1\over 6}(3+2a-4x_3)
  \left[a^2-4(1-x_3)(2-c-x_3)-2a(3+2x_3)\right]
\nonumber\\
  && -{1\over 2}\left(a(4-a)(2-c)+(1-c)\left[2(1-x_3)-a\right]^2\right)
  -{1\over{27}}(3+2a-4x_3)^3.
\end{eqnarray}
In Fig.~\ref{ps} we plot the full phase space limits and HERWIG's
limits.
Note that the collinear limit is $x_1=1$ and the soft limit is $x_3=0$
and that, as anticipated, HERWIG's missing region includes part of the
soft limit, which we avoid by using the cutoff
$x_3>2E_{\mathrm{min}}/m_t$.

In the dead zone we use the exact differential width, which, continuing
to neglect the $b$ mass, is:
\begin{eqnarray}
  \frac1{\Gamma_0}\frac{d^2\Gamma}{dx_1\,dx_3} &=&
  \frac1{(1-a)
  \left( 1 + \frac1a - 2a \right)}
  2 \;
  \frac{\alpha_s}{2\pi}
  \, C_F
  \frac1{x_3^2(1-x_1)}\Biggl\{\scriptstyle
 (1+\smfrac1a-2a)
 \left[(1-a)x_3-(1-x_1)(1-x_3)-x_3^2\right]
\nonumber\\&&\scriptstyle
 +(1+\smfrac1{2a})x_3(x_1+x_3-1)^2
 +2x_3^2(1-x_1)
 \Biggr\},
\end{eqnarray}
It is straightforward to use standard techniques to generate this
distribution within the allowed phase space region.

A final point involves the scale at which we evaluate $\alpha_s$.  As is
well known, using the transverse momentum of the emitted gluon sums up
an important class of next-to-leading logarithmic
corrections[\ref{CMW}].  However, we should be careful in using the
transverse momentum in the backward hemisphere, because in the $W$
direction it is zero, even though the gluon is not collinear to any
coloured parton.  We instead use the Durham-like generalization of
transverse momentum,
\begin{equation}
  \mu^2 = 2E_g^2(1-\cos\theta_{bg}) = \frac{x_3(1-x_1)}{2-x_1-x_3-a}m_t^2.
\end{equation}

\section{Soft Corrections}

According to Ref.~[\ref{S1}], the soft correction is implemented by
multiplying the emission probability for every emission {\em that is the
  hardest so far} by a correction factor that is simply the ratio of
HERWIG's differential distribution to the matrix element one.  The only
non-trivial part of this is in calculating the Jacobian factor
$J(x_1,x_3)$ of the transformation $(z,\xi)\to (x_1,x_3)$.  HERWIG's
distribution is then given by
\begin{equation}
  {{d^2\Gamma}\over{dx_1 dx_3}}={{d^2\Gamma}\over{dz d\xi}} \;
  J(x_1,x_3),
\end{equation}
where ${{d^2\Gamma}/{dz d\xi}}$ is given by the elementary emission
probability given in Eq.~(\ref{elementary}).  $J$~can be simply
calculated from the relations given earlier, but we have not found a
particularly compact form so we do not reproduce it here.

We recall that we only apply the soft correction for
$x_3>2E_{\mathrm{min}}/m_t$.

\section{Results}

The exact matrix element for $e^+e^- \to (bW^+)\,(\bar{b}W^-)\,g$ via
top quarks was calculated in Refs.~[\ref{OSS1},\ref{OSS2}] and compared
to HERWIG versions~5.8 and~5.9 respectively.  In fact, serious bugs have
been found in the treatment of top decays in both versions, so no firm
conclusion could be reached about the need for matrix element
corrections.

In this section, we repeat the analysis of Ref.~[\ref{OSS2}] in which
they consider a centre-of-mass energy only slightly above threshold,
$\sqrt{s}=360$~GeV, so that essentially all gluon emission is associated
with the top quark decays.  We work at the parton level (the final state
of the parton shower, without hadronization or $b$ decay) and cluster
all partons into three jets using the $k_\perp$ algorithm[\ref{CDOTW}].
Then the jets are required to be hard, ${E_T}_j>10$~GeV, and
well-separated\footnote{Note that the text of Ref.~[\ref{OSS2}] says
  that the angular cut is $\Delta R>0.4$, but it is clear from their
  Fig.~7(a) that it is actually $\Delta R>0.7$.}, $\Delta R>0.7$.  We
force the $W$ decays to be leptonic and do not include their decay
products in the jet clustering.

\begin{figure}
\vspace*{-1cm}
\begin{minipage}[t]{.47\linewidth}
  \centerline{
    \resizebox{!}{5.3cm}{\includegraphics{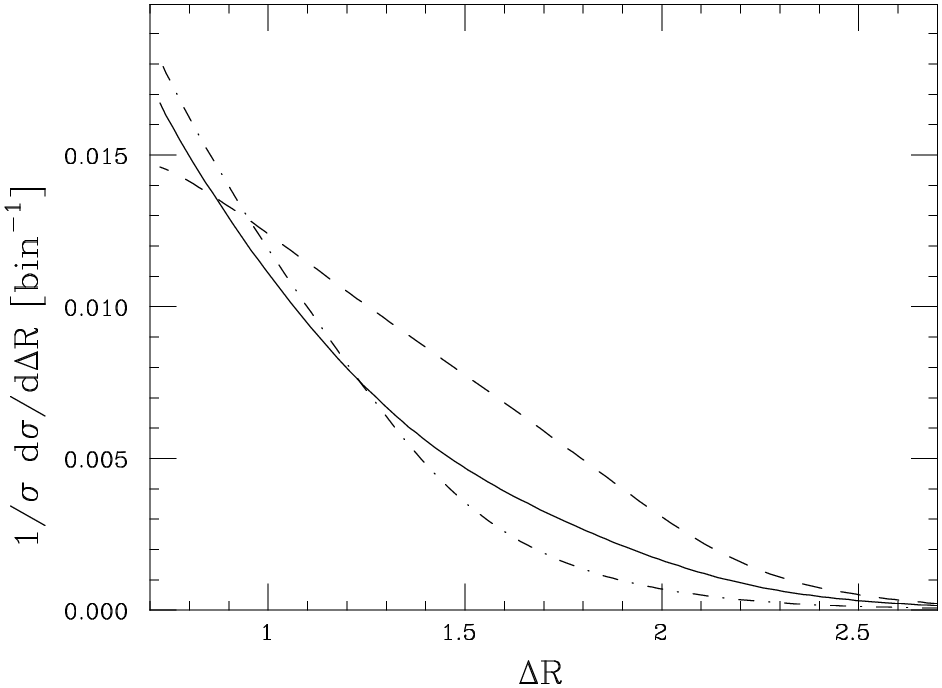}}
    }
  \caption[]{Differential distribution of invariant opening angle,
    $\Delta R^2=\Delta\eta^2+\Delta\phi^2$ for three-jet $e^+e^-\to
    t\bar{t}$ events at $\surd s=360$~GeV, according to HERWIG version
    5.9 (dashed), 6.0 (dot-dashed) and the version with matrix element
    corrections, 6.1 (solid).}
  \label{dr1}
\end{minipage}
\hspace{\fill}
\begin{minipage}[t]{.47\linewidth}
  \centerline{
    \resizebox{!}{5.3cm}{\includegraphics{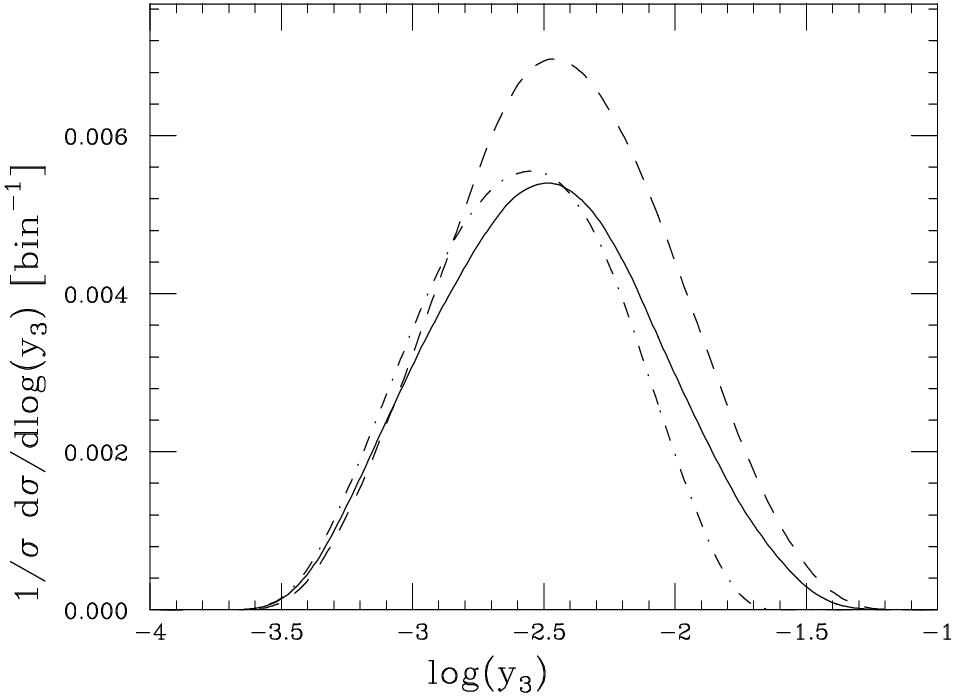}}
    }
  \caption[]{As Fig.~\protect\ref{dr1} but for the distribution of
    Durham jet algorithm measure,
    $y_3=\min_{ij}[\frac2s\min(E_i^2,E_j^2)(1-\cos\theta_{ij})]$.}
  \label{y1}
\end{minipage}
\end{figure}

In Figs.~\ref{dr1} and~\ref{y1}, we show the differential distribution
with respect to $\Delta R$, the separation of the closest pair of jets
in the event, and $\log y_3$, where $y_3$ is the value of the cutoff in the
$k_\perp$ algorithm at which the three jets would be merged to two,
according to the last public version,~5.9, version~6.0, which has the
bug that was found in~5.9 fixed, and a preliminary version that
includes our matrix element corrections,~6.1.
The version 5.9, which was used in Ref.~[\ref{OSS2}], clearly gives too
much gluon radiation.  This is easily explained by the bug which was
corrected in version 6.0\footnote{The wrong frame had been used for the
  $b$ parton shower: heavy quark decay was treated like all other
  processes and showered in the frame in which $E_b=\sqrt{p_t\cdot
    p_b}\approx110$~GeV instead of the top rest frame in which
  $E_b={{p_t\cdot p_b}\over{m_t}}\approx70$~GeV.}, which is seen to have
much less radiation at large angles and $y_3$ values.

The only difference for top decays between versions 6.0 and 6.1 is in
the matrix element correction discussed here.  Therefore Figs.~\ref{dr1}
and~\ref{y1} directly show its effect.  At large angle and $y_3$ the rate
of gluon emission is greatly enhanced due to the hard correction, which
fills the missing region of phase space.  At smaller angles and $y_3$
values, the soft correction gives a small reduction, showing that in the
uncorrected version the deficit at large angles was at least partially
compensated by a surfeit at small angles.

We turn now to a comparison with the tree-level matrix element results
of Ref.~[\ref{OSS2}].  In order to remove dependence on the electroweak
production process (for example the fact that HERWIG uses the cross
section for on-shell top quarks while the matrix element calculation
implicitly includes the top width) we compare the differential
distribution normalized to the total $e^+e^-\to t\bar{t}$ cross section.
Even after doing this, the normalization of a tree-level calculation is
still rather arbitrary, since the scale of $\alpha_s$ is not fixed.  We
choose to set $\alpha_s$ equal to its value at a typical transverse
momentum of jets passing the cuts, $\alpha_s(\approx\mbox{30
  GeV})\approx0.145$.
\begin{figure}
\vspace*{-1cm}
\begin{minipage}[t]{.47\linewidth}
  \centerline{
    \resizebox{!}{5.3cm}{\includegraphics{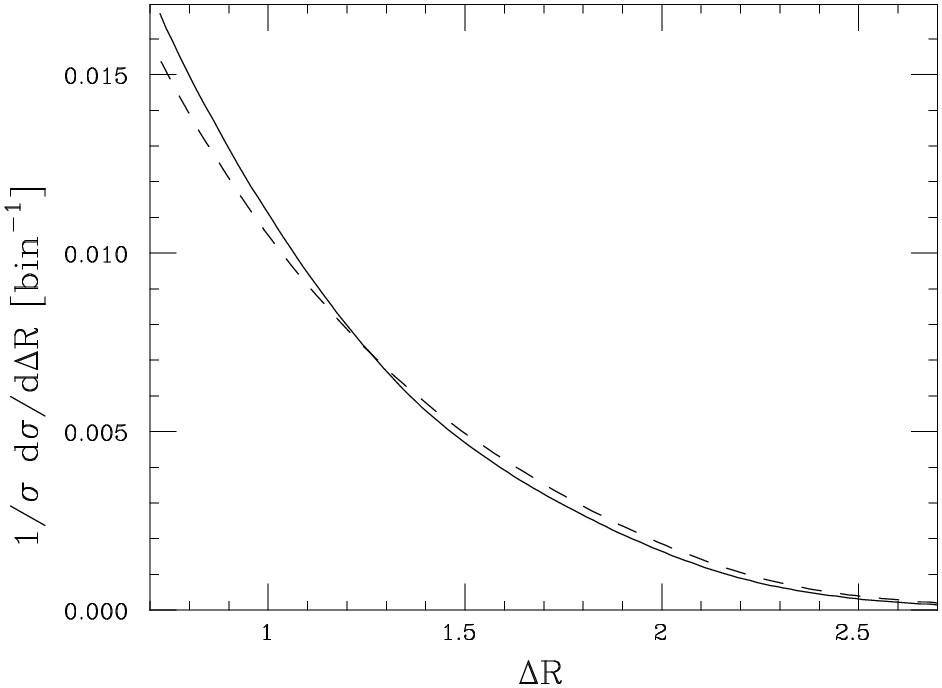}}
    }
  \caption[]{As Fig.~\protect\ref{dr1} but from HERWIG 6.1 (solid) and
    the tree-level calculation with $\alpha_s=0.145$ (dashed).}
  \label{dr2}
\end{minipage}
\hspace{\fill}
\begin{minipage}[t]{.47\linewidth}
  \centerline{
    \resizebox{!}{5.3cm}{\includegraphics{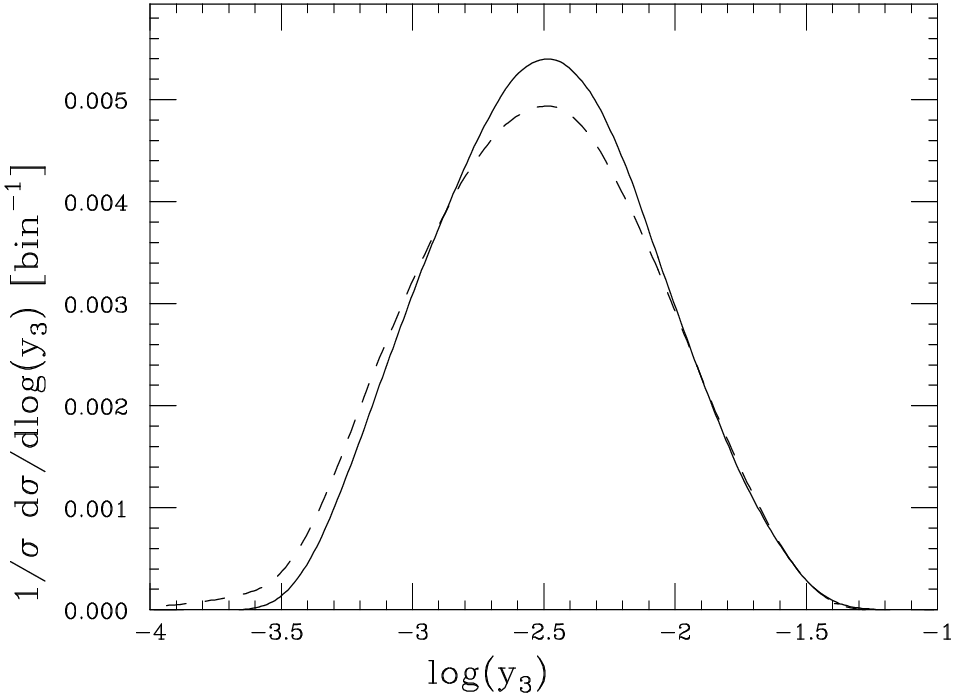}}
    }
  \caption[]{As Fig.~\protect\ref{y1} but from HERWIG 6.1 (solid) and
    the tree-level calculation with $\alpha_s=0.145$ (dashed).}
  \label{y2}
\end{minipage}
\end{figure}
Results are shown in Figs.~\ref{dr2} and~\ref{y2}.  After applying
matrix corrections the agreement is very good at large $y_3$ and $\Delta
R$.  As $y_3$ is reduced, we probe lower momentum scales, where $\alpha_s$
is getting larger in HERWIG, while it is fixed in the matrix element
calculation, so HERWIG lies above the calculation.  Finally, at very
small $y_3$, Sudakov suppression from multiple emission would be expected,
which is included in HERWIG but not in the matrix element calculation,
so HERWIG lies below the calculation.

\begin{figure}
\begin{minipage}[t]{.47\linewidth}
  \centerline{
    \resizebox{!}{5.3cm}{\includegraphics{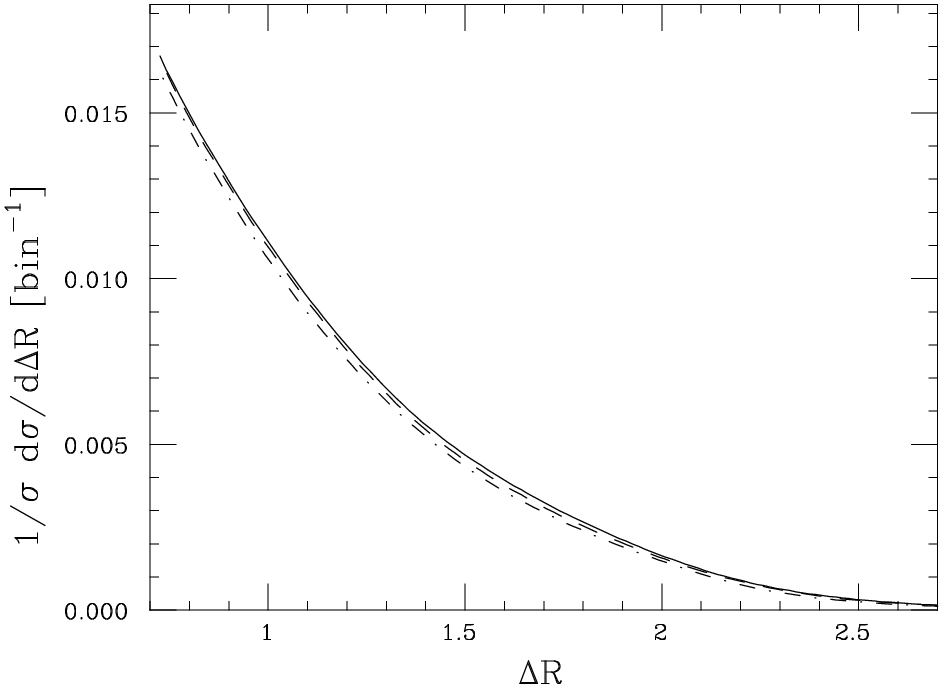}}
    }
  \caption[]{As Fig.~\protect\ref{dr1} but from HERWIG 6.1 with gluon
    energy cutoff set to 1~GeV (dashed), 2~GeV (solid) and 5~GeV
    (dot-dashed).}
  \label{dr3}
\end{minipage}
\hspace{\fill}
\begin{minipage}[t]{.47\linewidth}
  \centerline{
    \resizebox{!}{5.3cm}{\includegraphics{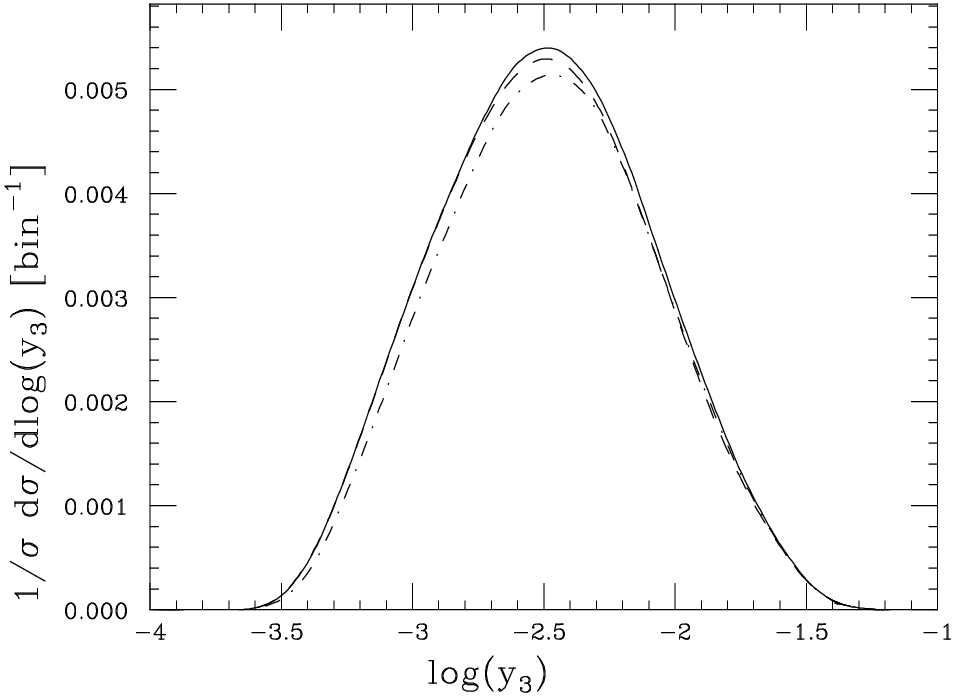}}
    }
  \caption[]{As Fig.~\protect\ref{y1} but from HERWIG 6.1 with gluon
    energy cutoff set to 1~GeV (dashed), 2~GeV (solid) and 5~GeV
    (dot-dashed).}
  \label{y3}
\end{minipage}
\end{figure}

Finally, in Figs.~\ref{dr3} and~\ref{y3} we show the dependence on the
arbitrary cutoff on soft gluon energies we introduced in the hard matrix
element correction.
It is clearly insignificant.  It is worth noting that the actual
fraction of events that have an emission in the dead zone generated by
the hard matrix element correction varies considerably with the cutoff,
from 2.4\% at 5~GeV to 3.8\% at 1~GeV, so the fact that the physical
distribution after cuts is unaffected is a non-trivial test of the
self-consistency of the procedure.  The default cutoff value, 2~GeV, was
used for the previous plots.

\section{Conclusions}

We have discussed HERWIG's parton shower treatment of top quark decay.
We applied the exact matrix-element result for emission in the region of
phase space that it does not populate and for the hardest emission in
the usually-populated region.  Since the missing region includes part of
the soft singularity, we have had to apply an arbitrary cutoff on the
energy of gluons emitted in this region.  The dependence on this cutoff
is however negligible after applying jet cuts.

HERWIG's predictions are considerably improved by this correction, and
agree well with the exact leading-order perturbative results.  Where
there are differences, they can be understood as advantages of the
parton shower approach.  We therefore feel confident that HERWIG now
gives a reliable treatment of top quark decays.

It is clearly interesting to ask how much these corrections affect the
top mass reconstruction from final state properties at the Tevatron.
However, before doing so it is essential to also implement matrix
element corrections to the top production processes.  This work is in
progress.

\section*{Acknowledgements}
We are grateful to Yuri Dokshitzer, Michelangelo Mangano, Pino
Marchesini, Stefano Moretti and Bryan Webber for fruitful discussions of
these and related topics and to Tim Stelzer for providing the computer
code to calculate the matrix element results in figures~\ref{dr2}
and~\ref{y2}.

\section*{References}
\begin{enumerate}
\item\label{FMNR}
  S. Frixione, M.L. Mangano, P. Nason and G. Ridolfi, {\it Heavy
    Quark Production}, to be published in {\it Heavy Flavours II},
  ed.\ A.J. Buras and M. Lindner, World Scientific [hep-ph/9702287]
\item\label{factorization}
  A. Bassetto, M. Ciafaloni and G. Marchesini, Phys.\ Rep.\ 100 (1983) 201 \\
  G. Altarelli and G. Parisi, Nucl.\ Phys.\ B126 (1977) 298
\item\label{MW1}
  G. Marchesini and B.R. Webber, Nucl.\ Phys.\ B238 (1984) 1
\item\label{HERWIG}
  G. Marchesini et al.\ Comput.\ Phys.\ Commun.\ 67 (1992) 465
\item\label{MW2}
  G. Marchesini and B.R. Webber, Nucl.\ Phys.\ B310 (1988) 461 \\
  R.K. Ellis, G. Marchesini and B.R. Webber, Nucl.\ Phys.\ B286 (1987)
  643; Erratum ibid.\ B294 (1987) 1180
\item\label{ADKT}
  Ya.I. Azimov, Yu.L. Dokshitzer, V.A. Khoze and S.I. Troyan, Phys.\
  Lett.\ B165 (1985) 147
\item\label{MW3}
  G. Marchesini and B.R. Webber, Nucl.\ Phys.\ B330 (1990) 261
\item\label{KS}
  V.A. Khoze and T. Sj\"ostrand Phys.\ Lett.\ B328 (1994) 466
\item\label{top}
  The CDF Collaboration (F. Abe et al.) Phys.\ Rev.\ Lett.\ 80 (1998)
  2779 \\
  The D0 Collaboration (B. Abbott et al.) Phys.\ Rev.\ D58 (1998) 052001
\item\label{S1}
  M.H. Seymour, Comput.\ Phys.\ Commun.\ 90 (1995) 95
\item\label{S2}
  M.H. Seymour, Z.\ Phys.\ C56 (1992) 161
\item\label{S3}
  M.H. Seymour, {\it Matrix Element Corrections to Parton Shower
    Simulation of Deep Inelastic Scattering}, contributed to 27th
  International Conference on High Energy Physics (ICHEP), Glasgow,
  1994, Lund preprint LU-TP-94-12, unpublished
\item\label{AS}
  J. Andr\'e and T. Sj\"ostrand, Phys.\ Rev.\ D57 (1998) 5767
\item\label{CMW}
  S. Catani, G. Marchesini and B.R. Webber, Nucl.\ Phys.\ B349 (1991) 635
\item\label{OSS1}
  L.H. Orr, T. Stelzer and W.J. Stirling, Phys.\ Lett.\ B354 (1995) 442
\item\label{OSS2}
  L.H. Orr, T. Stelzer and W.J. Stirling, Phys.\ Rev.\ D56 (1997) 446
\item\label{CDOTW}
  S. Catani, Yu.L. Dokshitzer, M. Olsson, G. Turnock and B.R. Webber,
  Phys.\ Lett.\ B269 (1991) 432
\end{enumerate}
\end{document}